\documentclass[prb,aps,amssymb,twocolumn,floatfix]{revtex4}

\usepackage{graphicx}% Include figure files
\usepackage{dcolumn}% Align table columns on decimal point
\usepackage{bm}% bold math

\begin{document}

\preprint{}

\title{Tuning alloy disorder in diluted magnetic semiconductors in high fields to 89 T}

\author{S. A. Crooker}

\affiliation{National High Magnetic Field Laboratory, Los Alamos
National Laboratory, Los Alamos, NM 87545}
\author{N. Samarth}
\affiliation{Department of Physics, Pennsylvania State University,  University Park, PA 16802}
\date{\today}
\begin{abstract}

Alloy disorder in II-VI diluted magnetic semiconductors (DMS) is typically reduced when the local magnetic spins align in an applied magnetic field.  An important and untested expectation of current models of alloy disorder, however, is that alloy fluctuations in many DMS compounds should increase again in very large magnetic fields of order 100 tesla.  Here we measure the disorder potential in a Zn$_{.70}$Cd$_{.22}$Mn$_{.08}$Se quantum well via the low temperature photoluminescence linewidth, using a new magnet system to $\sim$89 T. Above 70 T, the linewidth is observed to increase again, in accord with a simple model of alloy disorder.
\end{abstract}

%\pacs{75.50.Pp, 78.55.Et, 71.55.Gs, 78.66.Hf, 78.67.De}
\maketitle
%---------------------------------------------------------------
The linewidth of band-edge photoluminescence (PL) is a powerful and widely employed probe of disorder and crystal quality in compound semiconductors such as Al$_x$Ga$_{1-x}$As or Zn$_x$Cd$_{1-x}$Te.  Even the cleanest of these materials, however, exhibit compositional alloy disorder due to the random placement of cation (or anion) species. This intrinsic alloy disorder presents a microscopic fluctuation potential that can be directly inferred from the PL linewidth, $\Gamma$.  In PL studies, photoexcited excitons act as mesoscopic probes of compositional disorder, where each exciton averages  the alloy fluctuations over the atoms within its wavefunction.  The standard deviation of exciton energies gives the broadened PL linewidth.  $\Gamma$ can therefore be modeled by the statistics of alloy fluctuations over the $N$ cation sites within the exciton's nominal `volume'.

Models have been developed to quantify microscopic alloy fluctuations in compound semiconductors and to correlate this disorder with measurements of $\Gamma$. Initial work focused on non-magnetic semiconductors, in which alloy fluctuations are fixed for a given sample.\cite{Goede,Schubert,Mena}  More recently, alloy disorder in II-VI diluted magnetic semiconductors (DMS) has been studied.\cite{Ryabchenko,CrookerAnneal,Klar,Sugakov} In DMS the alloy disorder potential \emph{within a given sample} can be tuned as the magnetic spins (usually spin-$\frac{5}{2}$ Mn) align in an applied magnetic field, $H$.  DMS therefore provide a flexible system against which to benchmark models of alloy disorder. Experiments to date have shown a decrease in $\Gamma$ as Mn spins align in low $H$, and as Mn spin clusters begin to align in high $H$ to 60 T, in approximate agreement with models. However, a critical and as-yet-unverified aspect of current models is the expectation that alloy disorder in many DMS materials should \emph{increase} again as antiferromagnetically-bound clusters of Mn spins achieve full alignment at very high magnetic fields.\cite{CrookerAnneal}  This field scale is typically quite large, of order 50-100 T.

To this end we measure the PL linewidth ($\Gamma$) from a single Zn$_{.70}$Cd$_{.22}$Mn$_{.08}$Se quantum well to $\sim$89 T, using a new pulsed magnet system that allows repeated and non-destructive access to this ultrahigh field regime and -- importantly -- has sufficiently long pulse duration to permit collection of high-resolution PL data.  Above 70 T, we observe that $\Gamma$ markedly increases, in qualitative agreement with a simple `local-bandgap' model of compositional alloy disorder. These data further validate current models of field-tunable alloy disorder in DMS materials, and support `local-bandgap' approaches to alloy disorder in general.

Figure 1(a) depicts one such model of alloy disorder in DMS, using Zn$_{1-x}$Mn$_x$Se as an example. Shaded bars represent individual cation sites. Each cation site (either Mn or Zn, with probability $x$ and 1-$x$) is assigned the local bandgap of its parent compound, giving a fluctuating bandgap as a function of atomic position.  Owing to the large $J_{sp-d}$ exchange interaction in DMS,\cite{Furdyna} the local bandgap near a Mn cation is strongly dependent on its spin projection, $S_z$, which can assume values $-\frac{5}{2}, -\frac{3}{2},...,+\frac{5}{2}$. Thus, $E_{ZnSe}$=2.82 eV, and $E_{MnSe}$=3.4 eV $+ \frac{1}{2}(\alpha-\beta)S_z$ eV, where $(\alpha-\beta)$=1.37 eV is the $J_{sp-d}$ exchange integral in Zn$_{1-x}$Mn$_x$Se.

Statistically, some Mn cations have no magnetic neighbors.  These isolated, spin-$\frac{5}{2}$ Mn have Brillouin-like magnetization, $\langle S_z\rangle$=$-\frac{5}{2} B_{5/2}(5 \mu_B g H/2k_B T^*)$, where $T^*$ is an effective temperature that best fits the low-field susceptibility ($T^*$=3.5 K in this work). In addition, a calculable fraction of Mn cations will have exactly one Mn neighbor. These `Mn-Mn pairs' bind antiferromagnetically, with total spin $S_p$=0 at low fields.\cite{Furdyna,ShapiraReview}  Mn-Mn pairs align in step-wise fashion (giving $S_p$=1, 2, 3, 4, 5) whenever $H$ equals a multiple of their binding energy, of order 18 T in Zn$_{1-x}$Mn$_x$Se and 14 T in Zn$_{1-x}$Mn$_x$Te.

At $H$=0 the isolated Mn spins fluctuate, resulting in a large alloy disorder as depicted in Fig. 1(a).  $\Gamma$ can be modeled by the distribution of possible exciton energies, which in turn is computed from the statistics of alloy disorder. Following the approach of Ref. 5, an exciton has an energy $E$ given by the average bandgap of the $N$ cation sites within its effective `volume': $E=N^{-1}\sum^N_{j=1} e_j$. The average energy of all excitons, $\langle E \rangle$=$\mu$, is the mean bandgap and PL line center. The full-width at half-maximum PL linewidth, $\Gamma$, is then given by 2.355 times the standard deviation of exciton energies about $\mu$,
\begin{equation}
\langle (E- \mu)^2 \rangle^{1/2} = [(P_i \Delta_i^2 - \mu^2)/N]^{1/2},
\end{equation}
where $P_i$ is the probability that a particular cation is of type $i$ (\emph{e.g.}, a Zn cation, an isolated Mn with spin $S_z$, a paired Mn with spin $S_p/2$, etc), $\Delta_i$ is the local bandgap of that cation, and a sum over all cation types $i$ is implied.

At low temperatures and in modest fields (\emph{H}$<$10 T), all isolated Mn spins align, smoothing the alloy disorder\cite{Ryabchenko,CrookerAnneal,Klar,Sugakov} and reducing $\Gamma$. However, Mn-Mn pairs (having $S_p$=0 and local bandgap $E_p$=3.4 eV) still present a considerable alloy fluctuation, as drawn. At high $H$, $E_p$ drops in discrete steps as the pairs align and $S_p$=1,...,5. In Zn$_{.92}$Mn$_{.08}$Se, $E_p$ falls just below $\mu$ at the second step at $\sim$36 T (as drawn). At this point, the disorder potential is most smooth, and $\Gamma$ should attain its minimum value.  Fig. 1(b) shows $\Gamma$ calculated for the case of Zn$_{.92}$Mn$_{.08}$Se. This step-wise ``magnetic annealing" has been observed in 60 T fields in Zn$_{1-x-y}$Cd$_y$Mn$_x$Se quantum wells.\cite{CrookerAnneal}

\begin{figure}[tbp]
\includegraphics[width=.40\textwidth]{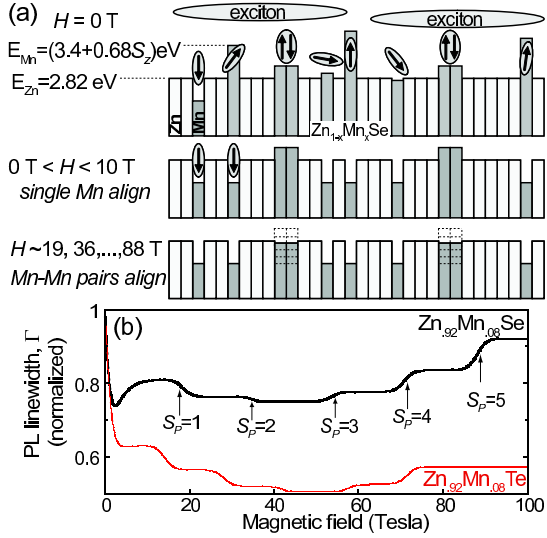}
\caption{(a) A model of alloy disorder in Zn$_{1-x}$Mn$_x$Se. Each cation, Zn or Mn, is given the bandgap of its parent compound (see text). At $H$=0, the isolated $S$=$\frac{5}{2}$ Mn spins fluctuate, and Mn-Mn pairs have net spin $S_p$=0.  At low $H$, isolated Mn align, reducing disorder. At high $H$, Mn-Mn pairs align in steps, reducing their local bandgap $E_p$. At the second step at 36 T, $E_p$ falls below the mean bandgap. At subsequent steps ($\sim$54, 71, 88 T), the net disorder is expected to increase. (b) $\Gamma$ calculated for Zn$_{.92}$Mn$_{.08}$Se (black), and Zn$_{.92}$Mn$_{.08}$Te (red).}
\label{fig1}
\end{figure}

However, a key prediction of this model is that alloy fluctuations can again \emph{increase} at even larger $H$ in DMS systems (such as Zn$_{1-x}$Mn$_x$Se and Zn$_{1-x}$Mn$_x$Te) where the bandgap of fully-polarized Mn-Mn pairs is less than $\mu$. As Fig. 1 depicts, further alignment of Mn-Mn pairs in Zn$_{.92}$Mn$_{.08}$Se at the third, fourth, and fifth magnetization steps reduces $E_p$ to values well below $\mu$, increasing the net alloy disorder -- and $\Gamma$ -- once again. These steps occur at $\sim$54, 71, and 88 T, and are therefore observable only in very large magnetic fields.

Figure 2 shows a schematic of the new `100 Tesla multi-shot magnet' \cite{Sims} at the National High Magnetic Field Laboratory. The three `outsert' coils are powered by a 1.4 GVA motor-generator that stores 650 MJ of available pulsed energy. In this work, the outsert provided a 2.5 s pulse with $\sim$36 T peak field.  The `insert' coil is separately powered by a 2.4 MJ capacitor bank (18 kV, 15.2 mF), providing up to an additional 54 T in a short pulse (5 ms rise, 8 ms decay). The present insert coil is rated to 90 T, to be replaced by a 100 T insert in 2007.  Figures 2(b,c) show the profile of a 88.9 T pulse, and an expanded view of eight pulses where the capacitor bank voltage was increased from 7-14 kV in 1 kV steps, giving peak fields from 63.9-88.9 T.  A pickup coil, calibrated against a de Haas-van Alphen Cu standard, measured $\partial H / \partial t$; integration errors give $\pm$0.5 T uncertainty in peak field. PL spectra were continuously acquired during the pulse using a 1.2 ms exposure time.

\begin{figure}[tbp]
\includegraphics[width=.37\textwidth]{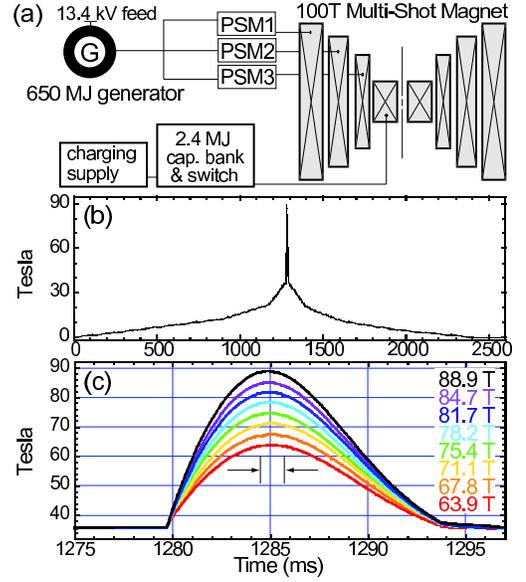}
\caption{(Color online) (a) The 100 Tesla multi-shot magnet. The outer coils are powered by a 1.4 GVA motor-generator and three power supply modules (PSM), and the `insert' coil is powered by a capacitor bank. (a) $H$ vs. time for a 88.9 T pulse. (c) An expanded profile of eight high-field pulses. Arrows denote the 1.2 ms PL acquisition time.} \label{fig2}
\end{figure}

We measure PL at 1.5 K from a 12 nm wide Zn$_{.70}$Cd$_{.22}$Mn$_{.08}$Se single quantum well with ZnSe barriers, grown by molecular beam epitaxy on (001)-oriented GaAs. The sample was excited with 800 $\mu$W of 405 nm laser light through a 600 $\mu$m diameter optical fiber.  PL was collected back through the same fiber and dispersed onto a CCD array in a 250 mm spectrometer. PL from this magnetic sample is entirely circularly ($\sigma^+$) polarized\cite{CrookerDMH} at low $H$, corresponding to recombination of the lowest energy, spin-down exciton.

Fig. 3(a) shows PL spectra at peak field for eight high-field pulses. At lower $H$  (0-53 T), over 2000 spectra were acquired in a separate pulse of a 60 T `Long-Pulse' magnet.\cite{CrookerAnneal,CrookerQ2D} The spectra have nearly gaussian lineshape, and exhibit a strong Zeeman redshift, $\Delta E_Z$, with increasing $H$.  $\Delta E_Z (H)$ tracks the average bandgap of the quantum well, providing a relative measure of the net Mn magnetization.\cite{Furdyna}  $\Delta E_Z (H)$ is shown in Fig. 3(b), using data from all high-field pulses (colored points) as well as 0-53 T data (black line). Following the rapid polarization of isolated Mn spins at low $H$, clear steps at $\sim$19, 36, and 53 T correspond to the first three magnetization steps of Mn-Mn pairs, as observed previously in this and related samples.\cite{CrookerAnneal,CrookerQ2D}  The fourth and fifth steps at $\sim$71 and 88 T are less clear; $\Delta E_Z$ increases more smoothly in this regime. The inset shows the total PL intensity.

Figure 3(c) shows the measured PL linewidth, $\Gamma$. Raw data is shown in black.  The points are derived from gaussian fits to the peak-field spectra of Fig. 3(a), and the solid line is derived from fitting all the low-field spectra ($<$53 T). The raw data includes a contribution to $\Gamma$ that increases nearly linearly with $H$.  This well-known phenomenon results \emph{only} from the field-induced shrinking of an exciton's `volume'.\cite{Mena} Smaller excitons are more sensitive to microscopic alloy fluctuations, giving a larger distribution of exciton energies and larger $\Gamma$.  A linear contribution to $\Gamma$ is therefore subtracted off, giving the data shown in red.  $\Gamma$ falls quickly at low $H$ ($<$5 T) as isolated Mn spins align and alloy fluctuations are smoothed.  $\Gamma$ then remains nearly constant until the first magnetization step of Mn-Mn pairs at $\sim$19 T.  At this step the net alloy disorder is reduced once again by the partial alignment of Mn-Mn pairs, where the local bandgap $E_p$ drops from 3.4 eV to a value closer to the mean bandgap $\mu$.  This `linewidth annealing' process repeats at the second and third steps. However, the high-field data show that this trend \emph{reverses} at the fourth magnetization step: above 70 T, $\Gamma$ unambiguously increases.  The error bars arise because $H$ is not perfectly constant during the 1.2 ms PL acquisition window near peak field.  If $H$=$H_0 \pm \delta H$ during a PL acquisition, the resulting Zeeman shift, $\partial E_Z/\partial H |_{H_0} \delta H$, acts to increase $\Gamma$. In this work, $\delta H$=0.8-2.4 T. For the low-field data, $\delta H$$<$0.15 T and errors in $\Gamma$ are negligible.

\begin{figure}[tbp]
\includegraphics[width=.40\textwidth]{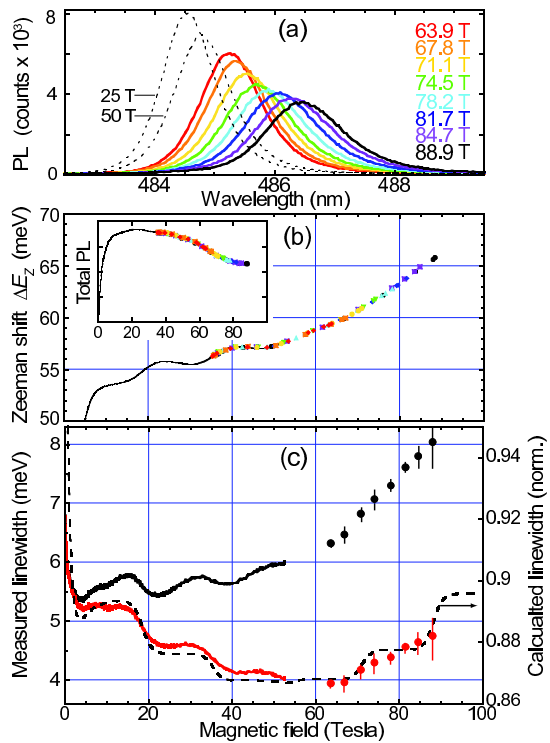}
\caption{(Color online) (a) PL spectra at 1.5 K from a Zn$_{.70}$Cd$_{.22}$Mn$_{.08}$Se quantum well. (b) The high-field PL Zeeman shift vs. $H$. Colored points (black line) denote data from the 100 T (60 T) magnet. Inset: total PL vs. $H$. (c) The measured linewidth, $\Gamma$ (black). Red data shows $\Gamma$ after subtraction of a linear term due to exciton shrinkage (see text). The dotted line shows the calculated, normalized linewidth.} \label{fig3}
\end{figure}

The dotted line in Fig. 3(c) shows $\Gamma$ calculated (and normalized) for Zn$_{.70}$Cd$_{.22}$Mn$_{.08}$Se.  $\Gamma$ should reach a minimum at the third step at 54 T, and increase thereafter. This behavior differs slightly from that predicted for Zn$_{.92}$Mn$_{.08}$Se in Fig. 1(b). The average bandgap $\mu$ in Zn$_{.70}$Cd$_{.22}$Mn$_{.08}$Se is smaller, such that $E_p$ falls just below $\mu$ at the third (rather than second) magnetization step. Thus, alloy disorder is not predicted to increase appreciably until the fourth step at 71 T, in clear agreement with the data.

These data therefore confirm a key high-field prediction of `local-bandgap' models of microscopic alloy fluctuations in DMS. However, although the measured stepwise decrease and subsequent increase of $\Gamma$ is qualitatively reproduced by the model, the overall magnitude of the effect is not. The measured changes in $\Gamma$ are larger than calculated: as seen in Fig. 3(c), the calculated linewidth is both scaled and offset.  A quantitative discrepancy is not, however, unexpected -- as formulated in Eq. 1, this model accounts only for alloy disorder as probed by excitons of \emph{fixed size}. In particular, the model does not attempt to account for any additional effects of exciton localization by alloy disorder.  Excitons can delocalize (increasing the number of cations within their volume, $N$) whenever disorder is smoothed, further reducing $\Gamma$ at these points. Klar \emph{et al.} \cite{Klar} have argued that delocalization of this sort must occur in DMS, based on $\Gamma (H)$ studies in Zn$_{.75}$Cd$_{.16}$Mn$_{.09}$Se.  We therefore suggest that a more complete theory may also account for the degree of delocalization that occurs when Mn spins align at low fields and at each magnetization step.\cite{Raikh}  As an indirect measure of localization and scattering by alloy disorder, note that the integrated PL emission (inset, Fig. 3b) roughly tracks the inverse of $\Gamma (H)$, suggesting that non-radiative recombination is suppressed when excitons delocalize in a smoother alloy potential.

We note finally that this simple model can also give a slight increase in $\Gamma$ at \emph{low} fields from 5-10 T due to the alignment of isolated Mn spins alone (\emph{e.g.}, dotted line in Fig. 3c). However, data from many DMS samples show only a weak effect.  We believe that the low-field magnetization of higher-order Mn clusters\cite{ShapiraReview} (triples, quadruplets, etc.), which are not explicitly included in this model, may obscure observation.

In summary, we use a newly-developed `100 T' magnet technology to experimentally confirm high-field predictions of a model for field-tunable alloy disorder in semiconductors. We gratefully acknowledge J. Sims and the magnet design group, and the 100 T commissioning team: C. Swenson, E. Serna, J. Schillig, P. Ruminer, D. Roybal, D. Rickel, A. Paris, M. Pacheco, J. Michel, J. Martin, M. Gordon, and J. Betts. This work was supported by the NSF, DOE and the State of Florida.
%---------------------------------------------------------------
%***************************************************************
%---------------------------------------------------------------

%\newpage
\end{document}